\documentclass{article}
\usepackage[psamsfonts]{amssymb}
\usepackage{amsmath}
\usepackage{cite}
\author{M. Alimohammadi$^1$\footnote{alimohmd@ut.ac.ir}\ \ and
M. Khorrami$^2$\footnote{mamwad@mailaps.org}
\\ $^1$ {\small School of Physics,
University of Tehran,}
\\ {\small North Karegar Ave., Tehran, Iran.}
\\ $^2$ {\small Department of Physics, Alzahra University, Tehran
1993891167, Iran.}}
\title{Phase transitions of Large-$N$ two-dimensional Yang-Mills and generalized Yang-Mills theories in the double scaling limit }
\date{}
\begin{document}
\maketitle
\begin{abstract}
\noindent The large-$N$ behavior of Yang-Mills and generalized
Yang-Mills theories in the double-scaling limit is investigated.
By the double-scaling limit, it is meant that the area of the
manifold on which the theory is defined, is itself a function of
$N$. It is shown that phase transitions of different orders occur,
depending on the functional dependence of the area on $N$. The
finite-size scalings of the system are also investigated.
Specifically, the dependence of the dominant representation on
$A$, for large but finite $N$ is determined.
\end{abstract}
\section{Introduction}
During recent years, Two-dimensional Yang-Mills theory (YM$_2$)
and generalized Yang-Mills theories (gYM$_2$'s) have extensively
been studied \cite{1,2,3,4,5,6,7,b,c,10,9,19,12,27,d,20}. These
are important integrable models which can shed light on some basic
features of pure QCD in four dimensions. Also, there exists an
equivalence between YM$_2$ and the string theory.

The starting point to make a correspondence between YM$_2$ and
string theory, is to study the large--$N$ limit of YM$_2$. For
example, as it is shown in \cite{3}, \cite{6}, and \cite{7}, a
gauge theory based on SU($N$) is split at large $N$ into two
copies of a chiral theory, which encapsulate the geometry of the
string maps. The chiral theory associated to the Yang--Mills
theory on a two--manifold $\cal M$ is a summation over maps from
the two--dimensional world sheet (of arbitrary genus) to the
manifold $\cal M$. This leads to a $1/N$ expansion for the
partition function and observables that is convergent for all of
the values of area$\times$coupling constant on the target space
$\cal M$, if the genus is one or greater.

Among the results obtained for YM$_2$, are the partition function
and the expectation values of the Wilson loops of YM$_2$ in
lattice- \cite{1,8} and continuum-formulations \cite{4,9,19}. The
partition function and the expectation values of the Wilson loops
of gYM$_2$'s have been calculated in \cite{c,10}. All of these
results are in terms of summations over the irreducible
representations of the corresponding gauge group. In general, it
is difficult to perform these summations explicitly and obtain
more closed results. However, for large groups these summations
are dominated by some specific representations in some cases, and
one can obtain closed-form expressions for that representation and
the observables of the theory.

In \cite{11}, the large-$N$ limit of the U$(N)$ YM$_2$ on a sphere
was studied. There the dominant (or classical) representation was
found and it was shown that the free energy of the U$(N)$ YM$_2$
on a sphere with the surface area $A<A_{\mathrm{c}}=\pi^2$ has a
logarithmic behavior \cite{11}. In \cite{12}, the free energy was
calculated for areas $A>\pi^2$, from which it was shown that the
YM$_2$ on a sphere has a third-order phase transition at the
critical area $A_{\mathrm{c}}=\pi^2$, like the well known
Gross-Witten-Wadia phase transition for the lattice two
dimensional multicolour gauge theory \cite{13,14}. The phase
structure of the large-$N$ YM$_2$, generalized YM$_2$'s, and
nonlocal YM$_2$'s on a sphere were further discussed in
\cite{27,20,37,39}. The large-$N$ limit of the partition function
of YM$_2$ on orientable compact surfaces with boundaries was
discussed in  \cite{AK}, and the large-$N$ behavior of Wilson
loops of YM$_2$ and gYM$_2$ on sphere were recently investigated
in \cite{l}. The critical behaviors of these quantities have been
also studied.

In \cite{AD}, U$(N)$-YM$_2$ theories were investigated, with the
property that the area of the manifold on which the theory is
defined, depends on $N$. It has been shown there that for a
specific parameterization of area (in terms of $N$), for which the
area tends to infinity as $N$ tends to infinity, there are
finite-size effects at large-$N$. By this it is meant that
although one expects that for infinite area, the dominant
representation be one for which the corresponding density is
everywhere one, for specific values of the parameters the dominant
representation is not that one. The difference between that
representation and the expected one, however, vanishes in terms of
the intensive quantities. That is, if one investigates the size of
the rows of the Young tableau themselves, there is a difference.
But if this size-difference is scaled by $N$ (divided by $N$) to
obtain an intensive quantity, it vanishes at the thermodynamic
limit. For the partition function too, a similar argument holds:
There is a difference between the logarithm of the partition
functions for the dominant representation and the expected
representation. But if one scales the logarithm of the partition
function with $N$ (divides it by $N^2$) to obtain intensive free
energy, then this difference vanishes at $N$ tends to infinity.

In this paper we want to study the large-$N$ behavior of YM$_2$
and gYM$_2$'s on a sphere, the area of which depends on $N$. In
section 2 some known results are reviewed, mainly to fix notation.
In section 3, the double scaling limit is introduced and various
kinds of the phase transitions are investigated. Here the emphasis
is on the intensive quantities, which remain finite at the
thermodynamic limit ($N\to\infty$). Any discontinuity in such
quantities in the thermodynamic limit is called a phase
transition. It is seen that such discontinuities are due to only
two factors: a discontinuity in the behavior of the area in the
thermodynamic limit, and a discontinuity in the behavior of the
free-energy density in terms of the area (the Douglas-Kazakov
phase transition). It is also shown that through the dependence of
the area on a parameter, one can to some extent control the order
of the transition. In section 4 a simple example is introduced and
the phases are explicitly studied for that example. Finally, in
section 5, the finite-size effects corresponding to the
double-scaling limit are investigated and the result of \cite{AD}
is generalized to the case of gYM$_2$'s.
\section{The classical representation}
Following \cite{10,20}, a generalized U$(N)$ Yang-Mills theory on
a surface is characterized by a function $\Lambda$:
\begin{equation}\label{w1}
\Lambda(R)=\sum_{k=1}^p\frac{a_k}{N^{k-1}}\,C_k(R),
\end{equation}
where $R$ denotes a representation of the group U$(N)$, $a_k$'s
are constants, and $C_k$'s are the Casimirs of the group defined
through
\begin{equation}\label{w2}
C_k=\sum_{i=1}^N[(n_i+N-i)^k-(N-i)^k].
\end{equation}
$n_i$'s are nonincreasing integers characterizing the
representation. It is assumed that $p$ is even and $a_p>0$. For
simplicity, from now on it is further assumed that all $a_k$'s
with odd $k$'s vanish. The partition function of such a theory on
a sphere of surface area $A$ is
\begin{align}\label{w3}
Z(A)&=\sum_R d_R^2\,\exp[-A\,\Lambda(R)],\nonumber\\
&=:\sum_R e^{S(R)},
\end{align}
where $d_R$ is the dimension of the representation $R$.

For large $N$, the summation on $R$ is dominated by the so called
classical representation, which maximizes the product
$d_R^2\,\exp[-A\,\Lambda(R)]$, as was shown in, for example,
\cite{20}. To obtain this representation, it is convenient to
introduce the new parameters
\begin{align}\label{w4}
x&:=\frac{i}{N},\nonumber\\
n(x)&:=\frac{n_i}{N},\nonumber\\
h(x)&:=-n(x)-1+x,
\end{align}
the density
\begin{equation}\label{w5}
\rho(h):=\frac{\mathrm{d} x}{\mathrm{d} h},
\end{equation}
and the function
\begin{equation}\label{w6}
G(z):=\sum_{k=0}^p a_k\,(-z)^k.
\end{equation}
Then $\rho_{\mathrm{cl}}$ (the density corresponding to the
classical representation) is characterized by
\begin{align}\label{w7}
\frac{A}{2}\,G(z)-\int\mathrm{d}
y\;\rho_{\mathrm{cl}}(y)\;\ln|z-y|&=\hbox{const.},\qquad
\mathrm{iff}\; \rho_{\mathrm{cl}}(z)\ne 1\hbox{
and } -a\leq z\leq a,\nonumber\\
\int_{-a}^a\mathrm{d} z\;\rho_{\mathrm{cl}}(z)&=1,
\end{align}
where $a$ is a positive number to be determined through the above
conditions.

For areas smaller than a critical area ($A_{\mathrm{c}}$), the
density corresponding to the classical representation is
everywhere less than one. For areas greater than $A_{\mathrm{c}}$,
there are places where $\rho_{\mathrm{cl}}$ is equal to one. And
for $A\to\infty$, the density tends to
\begin{equation}\label{w8}
\rho_{\infty}(y)=\begin{cases}
                                 1,& -\frac{1}{2}<y<\frac{1}{2}\\
                                 0,& \textrm{otherwise}
                    \end{cases}.
\end{equation}
The change of the behavior of $\rho_{\mathrm{cl}}$ at
$A=A_{\mathrm{c}}$ induces a phase transition, which is of third
order for the so-called typical theories, as discussed in \cite{l}
for example. That is, the free energy (density) of the system
defined as
\begin{equation}\label{w9}
F:=-\frac{1}{N^2}\,\log(Z),
\end{equation}
exhibits a discontinuous behavior at this area, and the
discontinuity is like $(A-A_{\mathrm{c}})^3$.
\section{The double scaling limit}
If the area $A$ is itself a function of $N$, then the phase
structure of the system may be different from what discussed
above. To be more specific, let's take $A$ to be a function of $N$
and some parameter $\alpha$, independent of $N$:
\begin{equation}\label{w10}
A=A(N,\alpha).
\end{equation}
The behavior of the system at $N\to\infty$, namely the density
$\rho$ corresponding to the classical representation and the free
energy $F$, is determined through the value of $A$ at
$N\to\infty$. So, defining
\begin{equation}\label{w11}
\mathcal{A}(\alpha):=\lim_{N\to\infty}A(N,\alpha),
\end{equation}
it is seen that at the thermodynamic limit ($N\to\infty$)
\begin{align}\label{w12}
F&=F(\mathcal{A}),\nonumber\\
\rho_{\mathrm{cl}}(z)&=\rho(\mathcal{A},z).
\end{align}
The dependence of $\mathcal{A}$ on $\alpha$, determines the phase
structure of the system. Let's assume that $\mathcal{A}$ is an
increasing function of $\alpha$. It may happen that $\mathcal{A}$
is a smooth function of $\alpha$. Then the system exhibits a third
order phase transition iff there exists an $\alpha_{\mathrm{c}}$
for which
\begin{equation}\label{w13}
\mathcal{A}(\alpha_{\mathrm{c}})=A_{\mathrm{c}}.
\end{equation}
For $\alpha$ greater than $\alpha_{\mathrm{c}}$, the system is in
the so called strong phase, where $\rho_{\mathrm{cl}}$ is equal to
one for some values of its argument, whereas for $\alpha$ less
than $\alpha_{\mathrm{c}}$, the system is in the so called weak
phase, where $\rho_{\mathrm{cl}}$ is everywhere less than one.

$\mathcal{A}$ may be a discontinuous function of $\alpha$ at some
$\alpha_{\mathrm{c}}$. In this case, obviously the free energy is
a discontinuous function of $\alpha$, and we have a zeroth order
transition at $\alpha_{\mathrm{c}}$. The discontinuity in
$\mathcal{A}$ may be such that for $\alpha$ greater than
$\alpha_{\mathrm{c}}$, $\mathcal{A}$ becomes infinite. In this
case, the free energy is not only discontinuous, but exhibits an
infinite jump at $\alpha_{\mathrm{c}}$, since the free energy
becomes infinite at infinite area. An example illustrating these,
is
\begin{equation}\label{w14}
A(N,\alpha)=\frac{\alpha}{\alpha_1}\,A_{\mathrm{c}}+b\,N^{\alpha-\alpha_2},
\end{equation}
where $b$, $\alpha_1$, and $\alpha_2$ are positive constants, and
$\alpha_1<\alpha_2$. One then has
\begin{equation}\label{w15}
\mathcal{A}(\alpha)=\begin{cases}\frac{\alpha}{\alpha_1}\,A_{\mathrm{c}},&\alpha<\alpha_2\\
                                 \infty,&\alpha>\alpha_2
\end{cases}.
\end{equation}
So, the system exhibits an infinite jump in the free energy at
$\alpha=\alpha_2$. Moreover, for $\alpha<\alpha_2$ the area
$\mathcal{A}$ is a smooth function of $\alpha$, and is less than
$A_{\mathrm{c}}$ for $\alpha<\alpha_1$ and greater than
$A_{\mathrm{c}}$ for $\alpha>\alpha_1$. So there is a third order
phase transition at $\alpha=\alpha_1$ as well. In terms of the
density corresponding to the classical representation:
\begin{itemize}
\item $\alpha<\alpha_1$. In this case $\max(\rho_{\mathrm{cl}})<1$,
that is the system is in the weak phase.
\item $\alpha_1<\alpha<\alpha_2$. In this case $\max(\rho_{\mathrm{cl}})=1$,
that is the system is in the strong phase.
\item $\alpha_2<\alpha$. In this case $\rho_{\mathrm{cl}}=\rho_\infty$,
that is the density is everywhere equal to one.
\end{itemize}

Another example is
\begin{equation}\label{w16}
A(N,\alpha)=A_{\mathrm{c}}+
\frac{b\,(\alpha-\alpha_{\mathrm{c}})^p\,N^{\alpha-\alpha_{\mathrm{c}}}
+c\,(\alpha-\alpha_{\mathrm{c}})\,N^{\alpha_{\mathrm{c}}-\alpha}
}{N^{\alpha-\alpha_{\mathrm{c}}}+N^{\alpha_{\mathrm{c}}-\alpha}},
\end{equation}
where $b$, $c$, and $p$ are positive constants. It is seen that
\begin{equation}\label{w17}
\mathcal{A}(\alpha)=\begin{cases}\mathcal{A}_w(\alpha):=A_{\mathrm{c}}+
c\,(\alpha-\alpha_{\mathrm{c}}),&\alpha<\alpha_{\mathrm{c}}\\
\mathcal{A}_s(\alpha):=A_{\mathrm{c}}+
b\,(\alpha-\alpha_{\mathrm{c}})^p,&\alpha>\alpha_{\mathrm{c}}
\end{cases}.
\end{equation}
So one obtains
\begin{equation}\label{w18}
F=\begin{cases}F_w[A_{\mathrm{c}}+c\,(\alpha-\alpha_{\mathrm{c}})],
&\alpha<\alpha_{\mathrm{c}}\\
F_s[A_{\mathrm{c}}+b\,(\alpha-\alpha_{\mathrm{c}})^p],
&\alpha>\alpha_{\mathrm{c}}
\end{cases}.
\end{equation}
$F_w$ and $F_s$ are the free energy in the weak and strong phase,
respectively. Noting that
\begin{equation}\label{w19}
F_s(A)-F_w(A)=q\,(A-A_{\mathrm{c}})^3+\cdots,
\end{equation}
where $q$ is a positive constant, one obtains
\begin{align}\label{w20}
F_s-F_w=&F_s[\mathcal{A}_s(\alpha)]-F_w[\mathcal{A}_w(\alpha)],\nonumber\\
=&q\,[\mathcal{A}_s(\alpha)-A_{\mathrm{c}}]^3+
F_w[\mathcal{A}_s(\alpha)]-F_w[\mathcal{A}_w(\alpha)]+\cdots,\nonumber\\
=&q\,b^3\,(\alpha-\alpha_{\mathrm{c}})^{3\,p}+
s\,[b\,(\alpha-\alpha_{\mathrm{c}})^p-c\,(\alpha-\alpha_{\mathrm{c}})]+\cdots,
\end{align}
where
\begin{equation}\label{w21}
s:=F'_w(A_{\mathrm{c}}).
\end{equation}
It is seen that if $p$ is less than one, then there is a phase
transition of the order $p$. If $p$ is greater than one, then
there is a transition of the order 1. If $p$ is equal to one, the
phase transition is of the order one (for $b\ne c$), or three (for
$b=c$).
\section{A simple example}
As a specific example, consider the following parameterization of
the area $A$:
\begin{equation}\label{w22}
A(N,\alpha)=\beta +\left(\frac{N}{2}\right)^\alpha,
\end{equation}
from which one obtains
\begin{equation}\label{w23}
\mathcal{A}=\begin{cases}
            \beta,& \alpha<0\\
            \infty,& \alpha>0
            \end{cases}.
\end{equation}
This is a simpler version of the first example in the previous
section. Simpler in the sense that for $\alpha<0$, $\mathcal{A}$
is constant and hence there is no third order phase transition.
There remains only a transition corresponding to an infinite jump
in the free energy.

If $\alpha<0$, the model is in the weak phase for
$\beta<A_{\mathrm{c}}=\pi^2$, and in the strong phase for
$\beta>A_{\mathrm{c}}$. In both cases, the density function
$\rho_{\mathrm{cl}}$ is not identical to one (for finite $\beta$),
that is, there is a value $y_0$ where $\rho_{\mathrm{cl}}(y)<1$ if
$|y|>|y_0|$. The precise value of $y_0$ depends on $\beta$. For
$\alpha>0$, the density $\rho_{\mathrm{cl}}$ is identical to one,
that is $\rho_{\mathrm{cl}}$ is equal to $\rho_\infty$.

Let's explicitly investigate the transition at $\alpha=0$. As for
$\alpha>0$ the area diverges, it is seen that the free energy also
diverges for $\alpha>0$. However, one can still compare the values
of the free energy for two different representations. To do so, we
follow the procedure introduced in \cite{AD}. For $\alpha>0$ where
$\rho_{\mathrm{cl}}$ is equal to $\rho_\infty$, the dominant
representation is the trivial representation and the parameters
corresponding to that are
\begin{equation}\label{w24}
n_i=i,\qquad -M\leq i \leq M.
\end{equation}
The parameter $i$ denotes the row of the Young Tableau and has
been taken between $-(N-1)/2$ and $(N-1)/2$, and
\begin{equation}\label{w25}
M:=\frac{N-1}{2}.
\end{equation}
For $\alpha<0$, the parameters corresponding to
$\rho_{\mathrm{cl}}$ are
\begin{equation}\label{w26}
n_i=\begin{cases}i,& |i|\leq M-l\\
i+\tilde n_i,& M-l<n_i\leq M\end{cases},
\end{equation}
where $\tilde n_i$ is a strictly increasing function of $i$. The
parameters $l$ and $\tilde n_i$ have to be determined by
maximizing the action in the $\alpha>0$ region. Denoting the
actions corresponding to the representations (\ref{w24}) and
(\ref{w26}) by $S_0$ and $S$, respectively, a calculation similar
to that performed in \cite{AD} results in
\begin{equation}\label{w27}
S-S_0=4\,l^2\,\left[\left(\ln\frac{M}{l}-\frac{A}{4}\right)\,
F_1+F_2\right].
\end{equation}
Defining
\begin{align}\label{w28}
u&:=\frac{i}{l},\nonumber\\
r(u)&:=\frac{\tilde n_i}{l},
\end{align}
for large $N$, the functions $F_1$ and $F_2$ can be written as
\begin{align}\label{w29}
F_1=&\int_0^1\mathrm{d}u\;r(u),\nonumber\\
F_2=&\int_0^1\mathrm{d}u\;\{r(u)\,[-\ln (u+r(u))+1-\ln
  2]+u\,[\ln u-\ln (u+r(u))]\}\nonumber\\
   &+\frac{1}{2}\,\int_0^1\mathrm{d}u\int_0^1\mathrm{d}v\;\ln\left[
   1+\frac{r(u)-r(v)}{u-v}\right].
\end{align}
To find the configuration which maximizes $(S-S_0)$, one puts
equal to zero the variations of $(S-S_0)$ with respect to $l$ and
$r$. Similar to \cite{AD}, these equations result in
\begin{equation}\label{w30}
\ln\left(\frac{M}{l}\right)-\frac{1}{4}\,M^\alpha+\frac{F_2}{F_1}
-\frac{\beta}{4}-\frac{1}{2}=0,
\end{equation}
and
\begin{align}\label{w31}
F_1=&\frac{1}{4},\nonumber\\
\frac{F_2}{F_1}=&\frac{1}{2}.
\end{align}
Note that our $F_2$ differs from one used in \cite{AD} by $-\beta
F_1/4$. Using (\ref{w30}) and (\ref{w31}), one obtains
\begin{align}\label{w32}
l=&M\,e^{-\frac{1}{4}\,(M^\alpha +\beta)},\nonumber\\
=&M\,e^{-\frac{A}{4}}.
\end{align}
Also, using (\ref{w27}), (\ref{w30}), and (\ref{w31}), one finds
\begin{align}\label{w33}
S-S_0=&\frac{l^2}{2},\nonumber\\
=&\frac{M^2}{2}\,e^{-\frac{1}{2}\,(M^\alpha
+\beta)},\nonumber\\
=&\frac{M^2}{2}\,e^{-\frac{A}{2}}.
\end{align}
If the partition function is dominated by one representation,
(\ref{w9}) can be rewritten as
\begin{equation}\label{w34}
F=-\frac{S}{N^2},
\end{equation}
from which,
\begin{align}\label{w35}
F-F_0=&-\frac{1}{8}\,e^{-\frac{1}{4}\,(M^\alpha
+\beta)},\nonumber\\
=&-\frac{1}{8}\,e^{-\frac{A}{2}},
\end{align}
or in the large-$N$ limit,
\begin{equation}\label{w36}
F-F_0=\begin{cases} -\frac{1}{8}\,e^{-\frac{\beta}{4}},&
\alpha<0\\
0,& \alpha>0
\end{cases}.
\end{equation}
This shows explicitly that for $\alpha>0$, the dominant
representation is the one corresponding to the density
$\rho_\infty$.

Calculations similar to this can be performed for $G(z)=z^k$-
gYM$_2$ model, with even $k$. To do so, one simply has to change
in $S$
\begin{equation}\label{w37}
-\frac{A}{2\,M}\,\sum_{i=1}^M n_i^2\to
-\frac{A}{(2\,M)^{k-1}}\,\sum_{i=1}^M n_i^k.
\end{equation}
One also has
\begin{equation}\label{w38}
n_i^k-i^k=k\,i^{k-1}\,\tilde n_i+\cdots,
\end{equation}
where only the leading term has been kept.

Using these, it is seen that for large $N$, the only change in the
results comes through
\begin{equation}\label{w39}
A\to A':=\frac{k}{2^{k-1}}\,A.
\end{equation}
It is then easy to see that (\ref{w36}) is changed to
\begin{equation}\label{w40}
F-F_0=\begin{cases} -\frac{1}{8}\,e^{-\frac{k\,\beta}{2^{k+1}}},&
\alpha<0\\
0,& \alpha>0
\end{cases}.
\end{equation}
Again we have a transition at $\alpha=0$.
\section{Finite-size effects}
Suppose that the parameterization of $A$ in terms of $\alpha$ is
such that for $\alpha>\alpha_{\mathrm{c}}$, the area $A$ is
infinite. It is then obvious that for
$\alpha>\alpha_{\mathrm{c}}$, and at the thermodynamic limit
$N\to\infty$, the dominant representation is that corresponding to
$\rho_\infty$. A question arises that is, if $N$ is large but not
infinite, is it still that representation which dominates the
partition function? The question may be rephrased like this. Take
another representation, for which the density $\rho$ is not
identical to one, and compare $S(R)$ in (\ref{w3}) for these two
representations.

This has been done in \cite{AD} for YM$_2$ and for the
parameterization
\begin{equation}\label{w41}
A=\alpha\,\log N+\beta.
\end{equation}
The result obtained there is that $\rho_{\mathrm{cl}}$ (the
density corresponding to the dominant representation) is equal to
one except for a narrow region around $(\pm 1/2)$, that is
\begin{equation}\label{w42}
\rho_{\mathrm{cl}}(z)\begin{cases} =1,& |z|\leq
\frac{1}{2}-\epsilon\\
<1,& |z|>\frac{1}{2}-\epsilon\end{cases},
\end{equation}
where
\begin{equation}\label{w43}
\epsilon\sim N^{-\alpha/4}.
\end{equation}
In terms of the parameters $n_i$ characterizing the representation
(eqs. (\ref{w25}) and (\ref{w26})), as calculated in \cite{AD}, it
turns out that for the dominant representation $R$ one has
\begin{align}\label{w44}
l&=e^{-\frac{\beta}{4}}\,M^{1-\frac{\alpha}{4}},\nonumber\\
&=M\,e^{-\frac{A}{4}},
\end{align}
and
\begin{align}\label{w45}
S-S_0&=\frac{l^2}{2},\nonumber\\
&=\frac{1}{2}\,e^{-\frac{\beta}{2}}\,M^{2-\frac{\alpha}{2}},\nonumber\\
&=\frac{M^2}{2}\,e^{-\frac{A}{2}}.
\end{align}

It is clear that for $\alpha<4$, both $l$ and $(S-S_0)$ diverge as
$N$ (or equivalently $M$) tends to infinity; whereas for
$\alpha>4$, both $l$ and $(S-S_0)$ tend to zero at the
thermodynamic limit. However, if one considers quantities properly
scaled by $N$ (so that they don't diverge at the thermodynamic
limit), one should investigate the behaviors of $(l/N)$ and $F$
(defined through (\ref{w9})) rather than those of $l$ and $S$. If
the partition function is dominated by a representation, then
(\ref{w9}) is rewritten like (\ref{w34}). Equations (\ref{w44})
and (\ref{w45}) are then transformed to
\begin{align}\label{w46}
\frac{l}{N}&=\frac{1}{2}\,e^{-\frac{\beta}{4}}\,M^{-\frac{\alpha}{4}},\nonumber\\
&=\frac{1}{2}\,e^{-\frac{A}{4}},
\end{align}
and
\begin{align}\label{w47}
F-F_0&=-\frac{1}{8}\,e^{-\frac{\beta}{2}}\,M^{-\frac{\alpha}{2}},\nonumber\\
&=-\frac{1}{8}\,e^{-\frac{A}{2}}.
\end{align}
It is obvious that in terms of these quantities, $\alpha=4$ is no
specific point. That is, $(l/N)$ and $(F-F_0)$ both tend to zero
as $N$ tends to infinity, as long as $\alpha$ is positive. This is
expected, as for positive $\alpha$ the area $A$ is infinite at the
thermodynamic limit, and for infinite area the dominant
representation is that corresponding to $\rho_\infty$.

Calculations similar to this can be performed for $z^k$-gYM$_2$
(with even $k$). To do so, one simply has to apply the changes
(\ref{w37}) and (\ref{w38}).

Using these, it is seen that for large $N$, the only change in the
results of \cite{AD} comes through (\ref{w39}). It is then obvious
that defining
\begin{align}\label{w48}
\alpha'&:=\frac{k}{2^{k-1}}\,\alpha,\nonumber\\
\beta'&:=\frac{k}{2^{k-1}}\,\beta,
\end{align}
one obtains for $l$ and $(S-S_0)$ (or $(l/N)$ and $(F-F_0)$)
results exactly similar to (\ref{w44}) and (\ref{w45}) (or
(\ref{w46}) and (\ref{w47})), but with $\alpha$ and $\beta$
replaced by $\alpha'$ and $\beta'$. So at the thermodynamic limit,
for $\alpha<(2^{k+1}/k)$ both $l$ and $(S-S_0)$ diverge, while for
$\alpha>(2^{k+1}/k)$ both $l$ and $(S-S_0)$ tend to zero. Also, at
the thermodynamic limit both $(l/N)$ and $(F-F_0)$ tend to zero
for any positive $\alpha$, as expected from the behavior of the
density corresponding to the dominant representation for large
areas.\\
\\
\textbf{Acknowledgement}:  M.A. would like to thank the research
council of the University of Tehran for partial financial support.

\end{document}